# Antiferromagnetism at $T > 500$ K in the Layered Hexagonal Ruthenate SrRu$_2$O$_6$

C. I. Hiley,[1] D. O. Scanlon,[2,3] A. A. Sokol,[2] S. M. Woodley,[2] A. M. Ganose,[2,3] S. Sangiao,[4] J. M. De Teresa,[4] P. Manuel,[5] D. D. Khalyavin,[5] M. Walker,[6] M. R. Lees[6] and R. I. Walton[1]*

*1. Department of Chemistry, University of Warwick, Gibbet Hill Road, Coventry, CV4 7AL, United Kingdom*

*2. University College London, Kathleen Lonsdale Materials Chemistry, Department of Chemistry, 20 Gordon Street, London, WC1H 0AJ, United Kingdom*

*3. Diamond Light Source Ltd., Diamond House, Harwell Science and Innovation Campus, Didcot, Oxfordshire OX11 0DE, United Kingdom*

*4. Insituto de Nanociencia de Aragón (INA), Universidad de Zaragoza, Zaragoza, Spain*

*5. ISIS Facility, Rutherford Appleton Laboratory, Harwell Science and Innovation Campus, Didcot, Oxfordshire, OX11 0QX, United Kingdom*

*6. Department of Physics, University of Warwick, Gibbet Hill Road, Coventry, CV4 7AL, United Kingdom*

*To whom correspondence should be addressed: E-mail:r.i.walton@warwick.ac.uk

We report an experimental and computational study of magnetic and electronic properties of the layered Ru(V) oxide SrRu$_2$O$_6$ (hexagonal, $P\bar{3}1m$), which shows antiferromagnetic order with a Néel temperature of 563(2) K, among the highest for 4$d$ oxides. Magnetic order occurs both within edge-shared octahedral sheets and between layers and is accompanied by anisotropic thermal expansivity that implies strong magnetoelastic coupling of Ru(V) centers. Electrical transport measurements using focused ion beam induced deposited contacts on a micron-scale crystallite as a function of temperature show $p$-type semiconductivity. The calculated electronic structure using hybrid density functional theory successfully accounts for the experimentally observed magnetic and electronic structure and Monte Carlo simulations reveals how strong intralayer as well as weaker interlayer interactions are a defining feature of the high temperature magnetic order in the material.

PACS codes: 75.50.Ee, 75.47.Lx , 71.15.Mb, 71.20.Nr

## I. INTRODUCTION

The diverse magnetic and electronic properties of the Sr–Ru–O system have been widely investigated, particularly for Ru(IV) oxides. $SrRuO_3$ is a rare example of a $4d$ ferromagnetic oxide [1], which is also a metallic conductor below its Curie temperature (160 K) [2]. $Sr_{n+1}Ru_nO_{3n+1}$ Ruddlesden-Popper phases include the spin-triplet chiral superconductor $Sr_2RuO_4$ [3], $Sr_3Ru_2O_7$, an itinerant metamagnet [4] with electron nematic behavior [5] and $Sr_4Ru_3O_{10}$, an itinerant ferromagnet and metamagnet [6,7]. While some strontium ruthenates contain Ru in the +5 or the +6 oxidation state, for example $Sr_2Ru_2^{V}Ru^{VI}O_{10}$ [8], $Sr_4Ru_2^{V}O_9$ [9] and $Sr_4Ru_2^{V}Ru^{VI}O_{12}$ [10], their magnetic properties have been less explored, despite the interesting phenomena observed in other Ru(V)-containing oxides [11-17]. The interest in ruthenates is part of a wider focus on the magnetism of $4d$ and $5d$ oxides, which differ considerably from the more widely studied $3d$ oxides when effects such as strong spin-orbit coupling are considered [18]. Recently, some of us reported the synthesis of some new ruthenium(V) oxides using solution chemistry, among which was the hitherto unreported $SrRu_2O_6$ [19]. This adopts the $PbSb_2O_6$-type structure, consisting of two oxygen layers in a hexagonal unit cell, with Ru occupying 2/3 of voids within one layer, while Sr occupies 1/3 voids in the second layer. Powder neutron diffraction data collected at room temperature on $SrRu_2O_6$ showed that the Ru(V) moments are antiferromagnetically ordered at room temperature in both the basal plane and along the principal axis, hexagonal type II, akin to $G$-type antiferromagnetic ordering in cubic lattices [20]. Herein we present powder neutron diffraction data from $SrRu_2O_6$ over the range 7.5 to 623 K to determine experimentally its Néel temperature, together with an investigation of electrical conductivity, and comprehensive calculations to examine the origin of electronic and magnetic properties.

## II. EXPERIMENTAL AND COMPUTATIONAL METHODS

A polycrystalline sample of $SrRu_2O_6$ was prepared hydrothermally at 200 °C [19]. Magnetisation was measured as a function of temperature using a Quantum Design MPMS-XL squid magnetometer with a furnace insert. Variable temperature powder neutron diffraction data were collected using the WISH diffractometer, ISIS, UK on heating from 7.5 to 623 K. The sample was loaded into a thin vanadium can (8 mm diameter), placed at the end of a stick which was placed into a closed cycle refrigerator (CCR). The end of the stick comprised a ceramic block, a copper block with sensors and heaters and the sample was surrounded by two heat shields, essentially providing a mini-furnace from the ceramic block.

For operation between 6 K and 300 K, the whole assembly was sat in exchange gas. For temperatures above room temperature, the CCR head was kept at 300 K, the exchange gas pumped out and the mini-furnace was used. Rietveld analysis was performed using the GSAS software [21]. To make resistivity measurements, a small amount of sample was suspended in ethanol and sonicated before being deposited onto a thermally oxidized silicon wafer containing pre-patterned metallic electrodes. A crystal was connected to the electrodes by focused ion beam induced deposition (FIBID) of Pt [22] and four-probe *in situ* measurements were made. An XPS spectrum was recorded using a Kratos Axis Ultra DLD spectrometer with the samples attached to electrically-conductive carbon tape, mounted on to a sample bar and studied at a base pressure of ∼ $2 \times 10^{-10}$ mbar at room temperature.

The electronic structure of $SrRu_2O_6$ was calculated using the HSE06 functional [23] as implemented in the VASP code [24,25]. The projector-augmented wave (PAW) [26] method was used to describe the interactions between the cores (Sr:[Kr], Ru:[Kr], and O:[He]) and the valence electrons. HSE06 has been shown to yield improved descriptions of structure, band gap and defect properties of a number of oxide semiconductors and transition metal oxides. Convergence with respect to k-point sampling and plane wave energy cut off were checked, and a cutoff of 750 eV and a k-point density of 0.2 k Å$^{-1}$ were found to be sufficient. Calculations were deemed to be converged when the forces on all the atoms were less than 0.01 eV Å$^{-1}$.

Magnetic ordering was explored by constructing hexagonal and orthorhombic supercells containing four Ru ions with differing nearest neighbor antiferromagnetic (AFM) and ferromagnetic (FM) spin alignments: AFM ordering (G-type); FM ordering (F-type); FM intralayer and AFM interlayer (A-type); AFM intralayer and FM interlayer (C-type); one AFM and two FM interactions intralayer for each Ru ("collinear AFM" within layers) and FM interlayer (labelled U-type); and two AFM and one FM interactions intralayer for each Ru and FM interlayer (labelled V-type). U-type can be visualized as FM stripes along the *a*-axis that are AFM connected intralayer and FM interlayer, whereas V-type are AFM stripes FM connected intra- and interlayer.

## III. RESULTS AND DISCUSSION

Figure 1 shows the magnetic susceptibility, $\chi$, recorded with temperature, $T$. Above room temperature the susceptibility increases with increasing temperature, with a discontinuity at ~570 K corresponding to the Néel temperature and a linear temperature dependence above $T_N$ up to the maximum temperature measured. A similar linear $\chi(T)$ behavior has been observed in the charge ordered antiferromagnet $Na_{0.5}CoO_2$ [27], in underdoped $La_{2-x}Sr_xCuO_4$ [28], and in several iron based superconductors [29,30] in the high temperature paramagnetic state. The susceptibility for $SrRu_2O_6$ at 400 K is much larger than the ~$10^{-4}$ emu mol$^{-1}$ observed in the 3$d$ $La_{2-x}Sr_xCuO_4$, but comparable with susceptibility observed in $Na_{0.5}CoO_2$ and the iron pnictides. In the last materials this is attributed to the coexistence of both local moments and itinerant electrons [29,30].

The decomposition temperature (673 K [19]) of $SrRu_2O_6$ precludes the preparation of a thermally densified pellet for transport property measurements and hence the conductivity of a single, micron-scale crystal selected from the polycrystalline sample was measured. The measured voltage shows a clear linear dependence with current, and a resistance of the crystal of 1.560(4) k$\Omega$ was obtained (Figure 2). By approximating the crystal shape and contact geometry to a cuboid (length, $l = 0.99$ μm, width, $w = 1.60$ μm, thickness, $t = 0.92$ μm) we estimate a resistivity, $\rho$, of $2.33 \times 10^5$ μ$\Omega$ cm. By an identical procedure, the $\rho$ of a second crystal was found to have a value on the same order of magnitude, $1.03 \times 10^5$ μ$\Omega$ cm. The resistivity as a function of temperature, measured from the first crystal, further confirms that $SrRu_2O_6$ is a semiconductor (Figure 2, inset). By using this Van der Pauw configuration [31] it was also possible to obtain a value for the Hall coefficient, $R_H$, of $6.098 \times 10^{-6}$ m$^3$ C$^{-1}$, indicative of a p-type semiconductor, with a hole density, $p$, of $1.025 \times 10^{18}$ cm$^{-3}$. In this configuration it is expected that the Hall voltage error is below 5% [32]. To understand these results, the electronic structure of $SrRu_2O_6$ was calculated. This confirmed the material to be semiconducting in nature, with a predicted band gap of 2.15 eV, Figure 3a. The valence band of the material displays pronounced O 2$p$ and Ru 4$d$ hybridization, while the conduction band minimum is dominated by unoccupied Ru 4$d$ states, mixed with some O 2$p$ states, Figure 3b. To compare our calculated results with experiment, we have overlaid simulated XPS data (constructed from the ion decomposed density of states weighted using the scattering cross sections of Yeh and Lindau [33]) over the experimental valence band spectrum in Figure 3c; the agreement between the two spectra corroborates the accuracy of our computational approach.

Figure 4 and Table I shows the results of Rietveld refinement of atomic and magnetic structure refinements against powder neutron diffraction data at temperature above and below the magnetic ordering temperature. The atomic structure is refined using the expected $P\bar{3}1m$ spacegroup for the PbSb$_2$O$_6$ structure, with no evidence for any structural phase transition over the temperature range studied, while the magnetic structure was solved using $P\bar{3}1c$ spacegroup, revealing an arrangement of spins such that there is antiferromagnetic order both within and between layers, Figure 5a inset, analogous to G-type order in cubic unit cell (hexagonal type II, according to the classification by Goodenough [20]). The ordered Ru$^{5+}$ moment in SrRu$_2$O$_6$, refined against *in situ* powder neutron diffraction data, was determined to be 1.425(10) μ$_B$ at 7.5 K [34]. This moment is significantly smaller than the spin-only value for a $d^3$ ion (3.87 μ$_B$), though is comparable to values obtained in other Ru$^{5+}$ ions in oxides [14, 17], which may be ascribed to some degree of covalency in M-O bonds in 4$d$ metal oxides [35]. The evolution of the ordered moment as a function of temperature, determined by Rietveld refinement, is shown in Figure 5a. The intensity of the magnetic Bragg peaks shows a loss in long-range magnetic ordering with a $T_N$ of ~ 570 K. Fitting with a power law [36] for the data above 440 K gives $T_N$ = 563(2) K. The ordering temperature is less than 100 K from the decomposition temperature of the material and unusually high for a 4$d$ oxide; to our knowledge the only 4$d$ oxide with a higher reported $T_N$ is the perovskite SrTcO$_3$, also containing a 4$d^3$ magnetic ion, with a value of ~1023 K [37].

**Table I**: Refined structural and magnetic Parameters for SrRu$_2$O$_6$ from Rietveld refinement of powder neutron diffraction data at two temperatures (see Figure 4 for Rietveld plots).

| Refined Parameter | $T$ = 7.5 K | $T$ = 623 K |
| --- | --- | --- |
| $a$ /Å | 5.20652(3) | 5.20586(3) |
| $c$ / Å | 5.22173(6) | 5.26186(6) |
| Scale factor | 399.0(6) | 359.7(5) |
| $R_p$ / % | 5.99 | 6.15 |
| $R_{wp}$ / % | 6.31 | 5.20 |
| $\chi^2$ | 15.56 | 10.66 |
| $U_{iso}$ (Sr)/ Å$^2$ | 0.0120(4) | 0.0280(5) |
| Ru Moment / $\mu_B$ | 1.43(1) | - |
| $U_{iso}$ (Ru)/ Å$^2$ | 0.0097(3) | 0.0190(3) |
| $x$(O)/$a$ | 0.3789(1) | 0.3791(1) |
| $z$(O)/$c$ | 0.2984(2) | 0.2982(2) |
| $U_{iso}$ (O)/ Å$^2$ | 0.0113(2) | 0.0257(2) |

Refinement of the lattice parameters of SrRu$_2$O$_6$ from the time-of-flight powder neutron diffraction data collected at temperatures from 7.5 to 623 K shows that whilst the *c*-axis increases with temperature, the *a*-axis displays a slight decrease from 5.20652(4) Å at 7.5 K to 5.20560(3) Å at 313 K, and remains constant from 313 K up to 623 K (Figure 5b). For comparison, we examined $SrSb_2O_6$ as a diamagnetic analogue with similar reduced mass: this displays linear thermal expansion in both the *a* and *c* axes from 100 to 648 K (Figure 5b). This anisotropic thermal expansivity of $SrRu_2O_6$ may be attributed to magnetoelastic coupling of the Ru(V) ions in the *a-b* plane, since the Ru – Ru intralayer distance remains constant over the whole temperature range studied. Since this behavior persists above $T_N$, it is possible that strong intralayer coupling on a local scale is present above the transition temperature which maintains the anisotropic expansivity.

The electronic structure calculations suggest that a distinct hybridization between the Ru and O states encourages electron transfer between Ru centers *via* O following a classical superexchange mechanism. Magnetic ordering was explored by constructing hexagonal and orthorhombic supercells containing four Ru ions (see Figures 6) with differing nearest neighbor antiferromagnetic (AFM) and ferromagnetic (FM) spin alignments. Crucially, the G-type AFM ordering found by neutron powder diffraction proved to be the most energetically favorable. The relative stabilities of the different spin configurations are given in Table II.

**Table II**: Relative energies of the different spin configurations tested for $SrRu_2O_6$. The alternative spin arrangements are all calculated at fully HSE06 relaxed geometries for each spin configuration, fixed geometries of the HSE06 relaxed G-type structure, and fixed to the experimentally reported structure.

| Configuration | Fully relaxed (eV) | HSE06 G-type structure (eV) | Expt Geometry (eV) |
|---|---|---|---|
| G | 0.000 | 0.000 | 0.000 |
| A | 0.457 | 0.602 | 0.624 |
| C | 0.002 | 0.002 | 0.002 |
| F | 0.461 | 0.614 | 0.638 |
| U | 0.138 | 0.194 | 0.202 |
| V | 0.281 | 0.385 | 0.404 |

Assuming a classical spin $^3/_2$ Heisenberg Hamiltonian description:

$$H = \sum_{k=1}^{4} J_k \sum_{\langle i,j \rangle^k} \underline{S}_i \cdot \underline{S}_j \tag{1}$$

where $\underline{S}_i$ is the spin on Ru atom $i$, and $J_k$ are the strengths of interaction between the $k^{th}$ nearest neighbors. In particular, $J_1$ is the strength of interaction between nearest neighboring Ru atoms within a honeycomb layer, which favors lattice antiferromagnetic behavior; $J_2$ involves second nearest neighbors within each hexagonal ring and opposes the effect of $J_1$; $J_3$ characterizes the interaction between honeycomb layers; and finally $J_4$ includes interactions between Ru atoms on opposite corners of hexagonal rings and reinforces the effect of $J_1$. Therefore, the model can be described as a parallel set of interacting "honeycomb $J_1$-$J_2$-$J_3$ systems" (2D hexagonal lattice of spins with $J_1$, $J_2$ and $J_4$ interactions) [38,39]. We have performed classical Monte Carlo (MC) simulations, as implemented within the program Spinner [40], of 11 honeycomb layers of 21 by 21 spins. 250,000 MC steps were performed, at each temperature after thermalizing, from 1200 K down to 0 K with a step of 5 K. The Spinner code uses energy units of $J_1$ and allows for a maximum of three types of interactions, therefore, a number of simulations were conducted for different combinations of couplings to establish the effect of $J_2/J_1$, $J_3/J_1$ and $J_4/J_1$.

We considered coupling constants for three sets of geometries: DFT optimized G-type spin configuration, fully optimized for each spin alignment, and experimental, obtained at room temperature. Using the differences between our calculated total hybrid density functional theory (DFT) energies for G-, F, A-, U- and V-type ordering (Table 1), we obtained $J_1 = 575$ K (428 and 599 K), $J_2/J_1 = 0.0291$ (0.0424 and 0.0275), $J_3/J_1 = 0.0287$ (0.0339 and 0.0330) and $J_4/J_1 = 0.0120$ (0.0169 and 0.0048) for G-type relaxed (fully relaxed and fixed) geometries. All interactions favor AFM between respective neighbors. Scaled by $J_1$ the phase transition between AFM G-type and a paramagnetic phase can be seen as a discontinuity in the slope of magnetic susceptibility with temperature, $d\chi/dT$. The statistical noise in our data made it difficult to determine accurately the critical temperature (Néel temperature, $T_N$). Therefore, the magnetic susceptibility simulated data have been filtered using moving averages over nine neighboring sample points, and we used the peak in the constant volume specific heat capacity ($C_v$) as a guide to where the discontinuity in $\chi$ lies; see Figure 7.

First we considered interaction only between nearest neighbors ($J_2=J_3=J_4=0$, dark blue curve in Figure 7) and found $T_N \sim 575$ K. Using only $J_1$ and $J_3$, we obtain $T_c = 0.5J_1$. Upon inclusion of interlayer interactions (non-zero $J_3$), which corresponds to a transition from two dimensional to three dimensional lattice of spins, we observed a significant enhancement of a peak in the $C_v$ curve accompanied by a shift of $T_N$ to higher temperatures by 64 K for $J_3/J_1$ = 0.0339, which yields a maximum critical temperature of $T_N$ = 639 K when $J_1$ = 575 K. Switching on additional intralayer interactions $J_2$ in turn weakens the AFM ordering and lowers $T_N$, whereas including $J_4$ interactions increases $T_N$.

The AFM ordering (exchange interaction) within the layers is much stronger than between them, consistent with the much shorter Ru – Ru distance in the layers. The small value of $J_3$ also explains the high stability (a low value of the relative energy) of the metastable C phase. The values of our calculated coupling constants, $J_k$, are also consistent with the ground state AFM arrangement both in classical and quantum mechanical phase diagrams for the "honeycomb $J_1$-$J_2$-$J_3$ system" [38,39]. Using the first set of coupling constants that characterize interactions in the DFT optimized G-type ordered phase, we estimate the Néel temperature to be 564 ± 5 K, very close to the experimentally measured value, whereas if we use the third set of coupling constants, obtained for the atomic structure that is fixed to that experimentally observed, then we estimate a slightly higher Néel temperature of 594 K.

Previous examples of high temperature magnetic ordering in 4$d$ and 5$d$ oxides have been restricted to materials with three-dimensional structures, such as the 5$d$ perovskites $SrTcO_3$ [37], $NaOsO_3$ [41] and $Sr_2CrOsO_6$ [42] (the last being ferromagnetic). Interestingly, three-dimensional antiferromagnetic order in the layered $PbSb_2O_6$ structure has also been seen for the materials $AAs_2O_6$, $A$ = Mn, Co, Ni and Pd [43,44], with $PdAs_2O_6$ showing a Néel temperature, $T_N$ of 140 K [45], but the magnetic ions here sit on the A-site of the structure, with greater interatomic separation from their magnetic neighbors than the B-site Ru in $SrRu_2O_6$. Our observation of high temperature, three-dimensional magnetic order in a two-dimensional, semiconducting material provides a new and structurally distinct system for further study of magnetism and electronic structure in 4$d$ oxides, with the possibility of preparation of doped or 5$d$ analogues of $SrRu_2O_6$. We note that during the preparation of this article Singh has published a theoretical study of $SrRu_2O_6$ using density functional theory [46], and has independently confirmed that magnetic anisotropy is high in this system, with comparable energy scales for moment formation and ordering which favours moments oriented along the $c$ axis.

We thank STFC for provision of beamtime at WISH. The work presented here made use of the ARCHER supercomputer through membership of the UK's HPC Materials Chemistry Consortium, which is funded by EPSRC grant EP/L000202. A.M.G. is grateful to Diamond Light Source for the partial funding of his studentship.

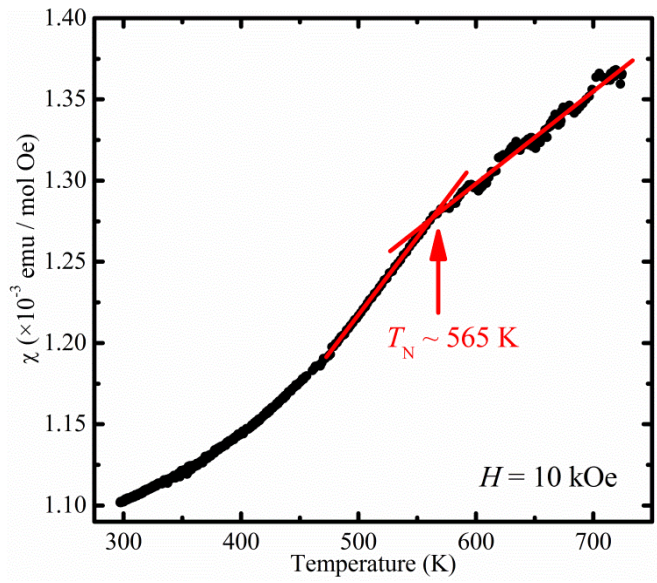

**Figure 1.** Magnetic susceptibility *vs* temperature for SrRu$_2$O$_6$ with the Néel temperature indicated. The red lines are linear interpolations of the data indicating a change in the slope of $\chi$(T) at the Néel temperature ~ 565 K.

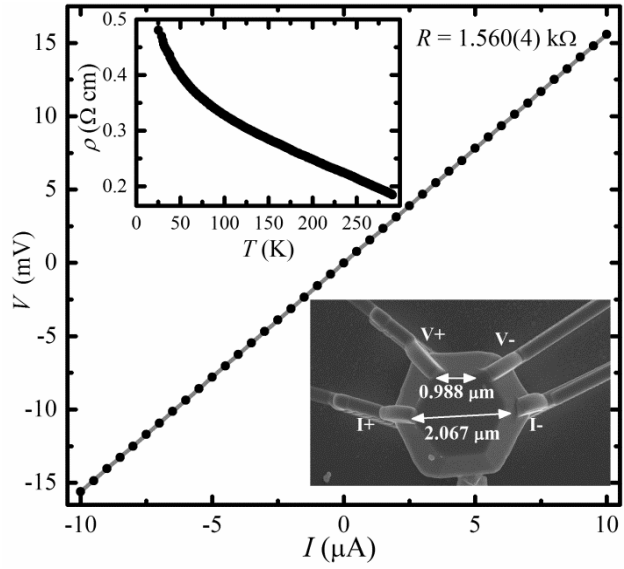

**Figure 2.** Voltage-current plot of a single crystal of $SrRu_2O_6$ (shown lower-right inset); experimental error bars are smaller than data points, but note that the absolute value of resistance has a large uncertainty due to the size determination of the crystallite. Resistivity as a function of temperature is plotted in the upper-left.

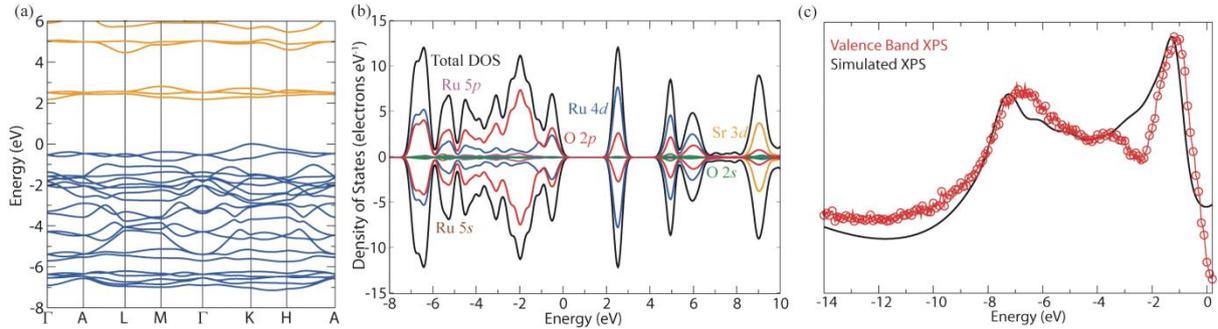

**Figure 3**. Calculated electronic structure of $SrRu_2O_6$ (a) Band structure, (b) density of states with partial contributions in two spin channels and (c) simulated XPS overlaid on the experimental XPS spectra. All were calculated using the G-type magnetic order found experimentally.

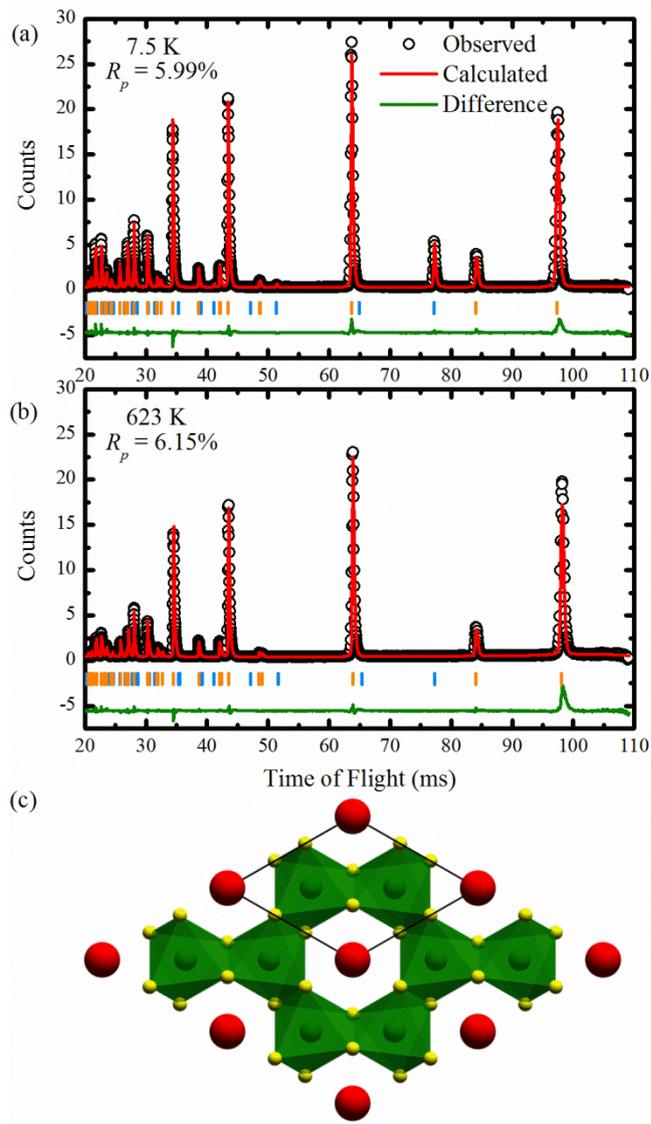

**Figure 4**. Rietveld fits of powder neutron data from SrRu$_2$O$_6$ (a) at 7.5 K and (b) at 623 K, and (c) a representation of the atomic structure of the material with green octahedra representing Ru and Sr shown as red spheres. In (a) and (b) blue ticks are due to the magnetic unit cell and orange due to the atomic unit cell.

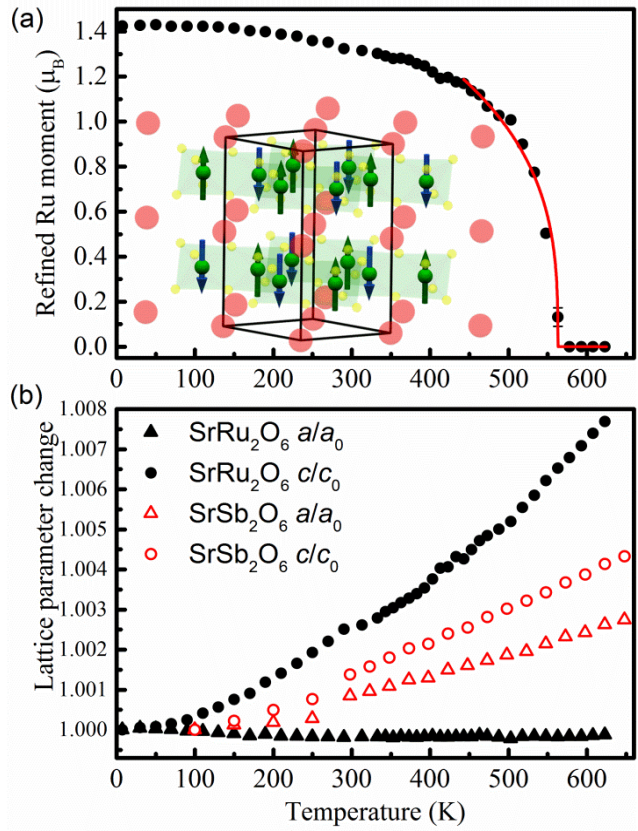

**Figure 5** (a) Ru(V) moment and (b) normalized lattice constants (divided by the lattice constants at 7.5 K) of $SrRu_2O_6$; as a function of temperature. In (a) the red line is the fitted power law from 430 K used to determined $T_N$. Lattice constants of non-magnetic $SrSb_2O_6$ as a function of temperature included (measured with laboratory powder X-ray diffraction and normalized to values at 10 K). Points on both plots without error bars have standard deviations smaller than the data points. Inset shows the magnetic cell, with red $Sr^{2+}$ ions, yellow $O^{2-}$ ions and antiferromagnetically ordered $Ru^{5+}$ ions in green.

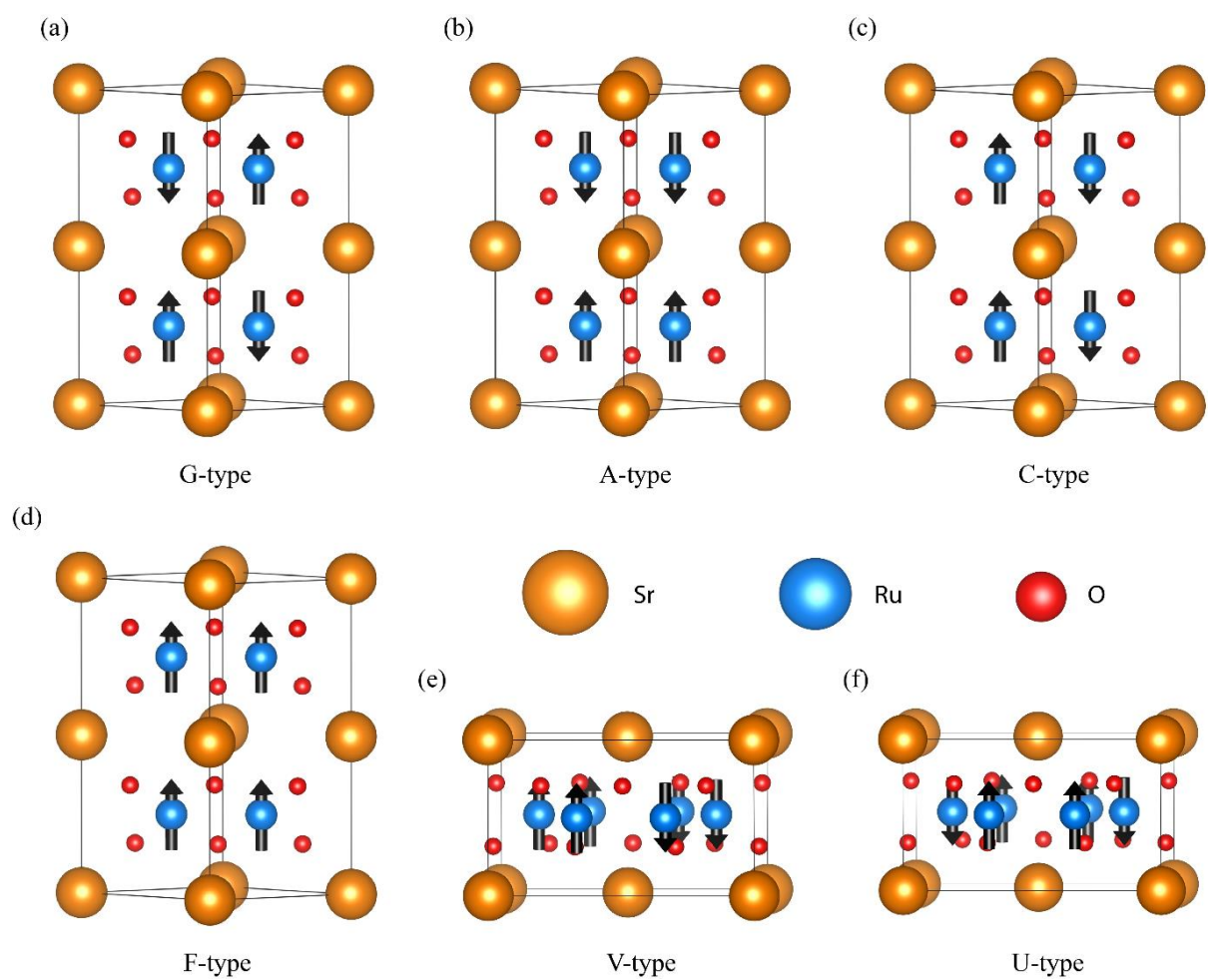

**Figure 6.** Spin configurations used in the DFT calculations (see text for explanation),

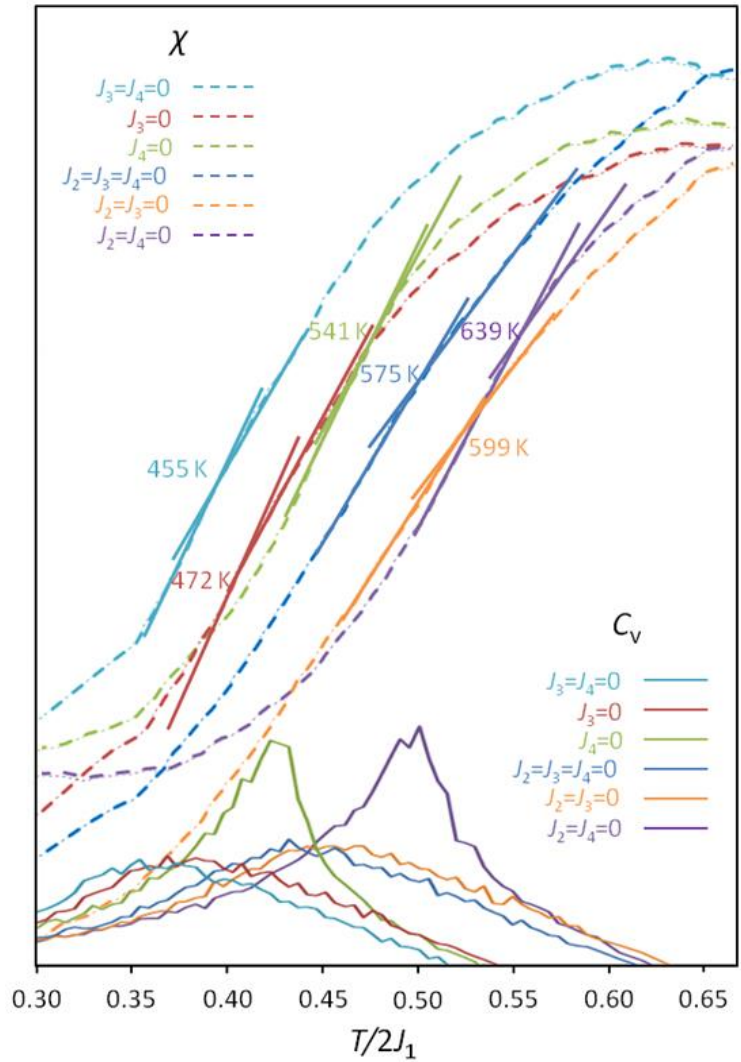

**Figure 7.** Magnetic susceptibility and constant volume heat capacity data obtained from Monte Carlo Simulations. The curves are obtained under different simplified models highlighted in the key and explained in the text. The factor of 2 in the scaling of temperature (energy) accounts for the double counting in the lattice sums.